\documentclass[preprint,showpacs,preprintnumbers,amsmath,amssymb]{revtex4}

\usepackage{graphicx}
\usepackage{dcolumn}
\usepackage{bm}
\usepackage[latin1]{inputenc}

\newcommand{\ud}[1]{{#1^{\dagger}}}

\newcommand\Tr{\mathrm{Tr}}

\begin{document}


\title{Strong-coupling of quantum dots in microcavities}

\author{Fabrice P.~Laussy}
\email{fabrice.laussy@uam.es}
\author{Elena del Valle}
\author{Carlos Tejedor}
\affiliation{%
Departamento de F\'{\i}sica Te\'orica de la Materia
Condensada, Universidad  Aut\'onoma de Madrid, Spain
}%

\date{\today}

\begin{abstract}
  We show that strong-coupling (SC) of light and matter as it is
  realized with quantum dots (QDs) in microcavities differs
  substantially from the paradigm of atoms in optical cavities.  The
  type of pumping used in semiconductors yields new criteria to
  achieve SC, with situations where the pump hinders, or on the
  contrary, favours it.  We analyze one of the seminal experimental
  observation of SC of a QD in a pillar microcavity [Reithmaier
  \emph{et al.}, Nature (2004)] as an illustration of our main
  statements.
\end{abstract}

\pacs{42.50.Ct, 78.67.Hc, 42.55.Sa, 32.70.Jz}
\maketitle

\begin{figure}[thbp]
  \centering
  \includegraphics[width=.66\linewidth]{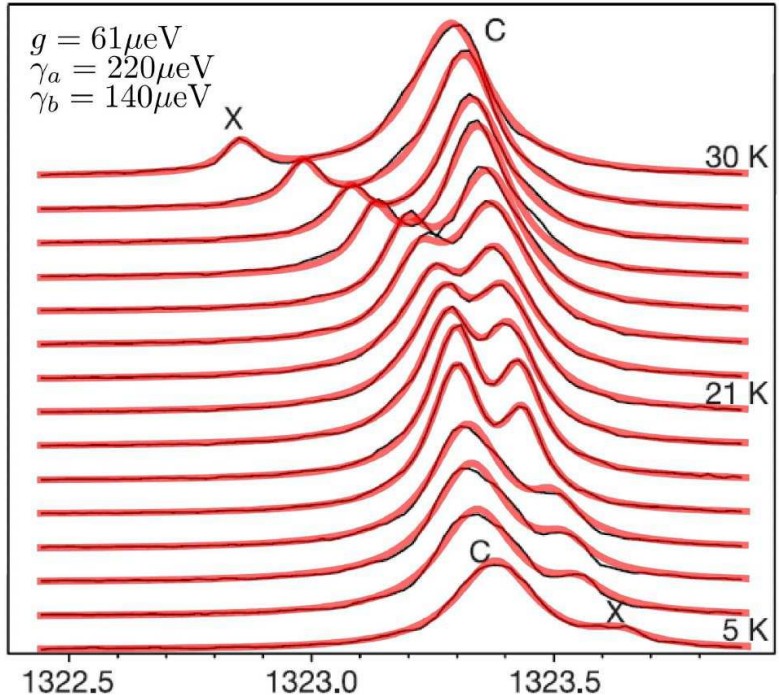}
  \caption{Anticrossing of the the cavity (C) and exciton (X)
    photoluminescence lines as reported by Reithmaier \emph{et al.} in
    Nature (2004), demonstrating SC in their system. Energies are
    given in meV. The red lines are our superimposed fits with the
    best global fit parameters in the top left corner. Such a good
    agreement cannot be obtained neglecting the pump-induced
    decoherence.}
  \label{ThuMar6011242UTC2008}
\end{figure}

The so-called \emph{strong-coupling} (SC) regime occurs in a coupled
system where the interaction strength overcomes the losses.  A case of
fundamental interest is that of light (photons) and matter (atoms,
electrons, etc.), coupled by the electromagnetic force.  That this
coupling is usually so weak accounts for the tremendous success of
quantum electrodynamics (QED), that affords all the required accuracy
at the level of perturbation theory. By confining the emitter in a
cavity, repeated interactions with the trapped photon(s) occur and
strong-coupling can thus be obtained, as was demonstrated in
pioneering experiments with Rydberg atoms migrating in optical
cavities~\cite{thompson92a}. Every achievement with atoms becomes an
objective for semiconductors, that offer unique advantages in terms of
integration and scalability, but come with their disadvantages in the
form of the overall complexity brought by condensed matter over its
fundamental elements.  Strong-coupling was first reported in
semiconductors with quantum wells in planar
microcavities~\cite{weisbuch92a}, that launched a new field
investigating exotic phases of matter in condensed systems such as
Bose-Einstein condensates~\cite{kasprzak06a} and
superfluids~\cite{amo07a_psip}. There is not yet a general consensus
that quantum physics rules these systems and that nonlinear optics
could not equally or better account for the observed phenomena.  A
more exact analogue to the atomic case that sticks closer to the
quantum regime is provided by zero dimensional (0D)
heterostructures---quantum dots (QDs)---where the material
excitations---the excitons---are fully quantised. SC in such systems
was only recently realized~\cite{reithmaier04a,yoshie04a,peter05a}.
Fig.~\ref{ThuMar6011242UTC2008} reproduces one of these seminal
contributions, by Reithmaier \emph{et al.}~\cite{reithmaier04a}, with
QDs in micro-pillars.
The most striking feature of strong coupling is the splitting of the
spectral shape when the system is at resonance: the line of the cavity
and that of the emitter, both at the same frequency, do not
superimpose but anticross with a splitting related to the coupling
strength. The doublet observed in Fig.~\ref{ThuMar6011242UTC2008} at
21K, when the two modes are expected to be resonant, demonstrates that
they are strongly-coupled, and this observation is the central result
of Ref.~\cite{reithmaier04a} and of the related
works~\cite{yoshie04a,peter05a}. In the ample literature devoted to
the description of this spectral shape~\cite{sanchezmondragon83a,
  agarwal86a, carmichael89a, savona94a, andreani99a}, the seminal work
of Ref.~\cite{sanchezmondragon83a} paid little attention to the
excitation (using a coherent state as an initial condition) and
neglected dissipation, while works such as~\cite{agarwal86a,
  savona94a} addressed the case of coherent pumping.
Ref.~\cite{carmichael89a} described spontaneous emission of an excited
state, and thus the initial state was fixed to be the excited state of
the atom in an empty cavity. This was repeated in the theoretical work
addressing the semiconductor case~\cite{andreani99a}, where, however,
a more complicated dynamics enters the picture. In practice, it is not
possible to initialise the cavity-emitter system in a semiconductor as
it is in the atomic case, where atoms can be singled out, manipulated
and sent one by one into the cavity.  Semiconductor QDs in a cavity
are typically excited far above resonance and electron-hole pairs
relax incoherently to excite the QD in a continuous flow of
excitations, establishing a steady state that washes out the coherent
Rabi oscillations.  Therefore, a Fock state as an initial state does
not correspond to the experimental reality.  Instead, the system is
maintained in a mixed state with probabilities~$p(n)$ to realize
the~$n$th excited state, that are imposed by the experimental
configuration.

In this Letter, we provide the appropriate theoretical model to
describe SC with 0D semiconductors. For the sake of illustration, we
support our discussion with the results of
Fig.~\ref{ThuMar6011242UTC2008}, that our model reproduces with an
excellent agreement. On the contrary, other models, with their
particular initial condition~\cite{sanchezmondragon83a, carmichael89a,
  andreani99a} cannot account for these spectra beyond the mere
prediction of the line splitting.
The shortcoming of downplaying the importance of the quantum state
that is realized in the system owing to pumping, has as its worst
consequence a misunderstanding of the results,
the most likely being the qualification of weak-coupling (WC) for a
system in SC that cannot be spectrally resolved because of
decoherence-induced broadening of the lines.  Being blind to the
theory makes the track for SC in 0D semiconductors particularly
difficult, involving a strong element of chance.  Understanding the
excitation scheme drastically reduces this element of hazard, as we
shall see below. Most importantly, our model unravels the physics
behind the experimental result, by spelling out which quantum state
has been produced, by providing most-likelihood estimators of the
sample parameters, by distinguishing the Bose of Fermi-like character
of its excitations and by predicting results as the excitation is
changed, most interestingly, as the pump is increased and the system
is brought into the nonlinear regime.

We describe the system with a quantum dissipative master equation
$\partial_t\rho=\mathcal{L}\rho$ for the density matrix
$\rho$~\cite{carmichael02a}, with the Liouvillian~$\mathcal{L}$
defined by its action on any operator~$O$ of the tensor product of the
light and matter Hilbert spaces~$\mathcal{H}_a$ and~$\mathcal{H}_b$:
\begin{align}
  \label{eq:ThuOct18162449UTC2007}
  &\mathcal{L}O=i[O,\omega_a\ud{a}a+\omega_b\ud{b}b+g(\ud{a}b+a\ud{b})]\\
  &+\sum_{c=a,b}\Big(\frac{\gamma_c}2(cO\ud{c}-\ud{c}cO)+\frac{P_c}2(\ud{c}Oc-c\ud{c}O)+\mathrm{h.c.}\Big)\,,\nonumber
\end{align}
where $g$ is the interaction strength between the cavity mode---with
annihilation operator~$a$ at energy~$\omega_a$---and the material
excitation---with operator~$b$ at energy~$\omega_b$---and respective
decay and pumping rates~$\gamma_{a,b}$ and $P_{a,b}$.  An important
experimental parameter is the detuning between the bare modes,
$\Delta=\omega_a-\omega_b$, that can be tuned effectively by changing
the temperature. In our case where the interplay of pumping and
dissipation establishes a steady state, the system is ergodic and the
cavity emission spectrum follows from the Wiener-Khintchine theorem as
$S(\omega)\propto\lim_{t\rightarrow\infty}\Re\int_0^\infty\langle\ud{a}(t)a(t+\tau)\rangle
e^{i\omega\tau}d\tau$.
According to the quantum regression theorem, a set of
operators~$A_{\{\alpha\}}$ that
satisfy~$\Tr(A_{\{\alpha\}}\mathcal{L}O)=\sum_{\{\beta\}}
M_{\{\alpha\beta\}}\Tr(A_{\{\beta\}}O)$ for
all~$O\in\mathcal{H}_a\otimes\mathcal{H}_b$ for
some~$M_{\{\alpha\beta\}}$, yields the equations of motion for the
two-time correlators as $\partial_\tau\langle
O(t)A_{\{\alpha\}}(t+\tau)\rangle=\sum_{\{\beta\}}M_{\{\alpha\beta\}}\langle
O(t)A_{\{\alpha\}}(t+\tau)\rangle$.  If~$b$ is a Bose operator
like~$a$ (the photon operator), $M$ is defined
by~$M_{\substack{nm\\nm}}=-i(n\omega_a+m\omega_b)-n\Gamma_a/2-m\Gamma_b/2$,
$M_{\substack{nm\\n+1,m-1}}=M_{\substack{mn\\m-1,n+1}}=-igm$ and zero
otherwise, where we defined as a shortcut the \emph{effective
  broadenings}~$\Gamma_{a,b}=\gamma_{a,b}-P_{a,b}$.  We also introduce
the following notation:
\begin{equation}
  \label{eq:ThuMar6130200UTC2008}
  \gamma_\pm=(\gamma_a\pm\gamma_b)/4\quad\mathrm{and}\quad\Gamma_\pm=(\Gamma_a\pm\Gamma_b)/4\,.
\end{equation}
Instead of \emph{ad hoc} initial conditions for the cavity population
and off-diagonal coherence, such as those provided by the excited
state of the QD~\cite{carmichael89a,andreani99a}, we use the steady
state values obtained by solving~$\Tr(\ud{a}a\mathcal{L}\rho)=0$
and~$\Tr(\ud{a}b\mathcal{L}\rho)=0$. For instance, the steady state
cavity population
$n_a=\lim_{t\rightarrow\infty}\langle\ud{a}a\rangle(t)$, reads:
\begin{equation}
  \label{eq:FriOct26003321UTC2007}
  n_a=\frac{g^2\Gamma_+(P_a+P_b)+P_a\Gamma_b(\Gamma_+^2+(\frac{\Delta}{2})^2)}{4g^2\Gamma_+^2+\Gamma_a\Gamma_b(\Gamma_+^2+(\frac{\Delta}{2})^2)}\,.
\end{equation}

The equations are closed and the normalised photoluminescence (PL)
spectrum $S(\omega)$ can therefore be expressed analytically:
\begin{equation}
  \label{eq:ThuMar6001103UTC2008}
  S(\omega)=(\mathcal{L}_1+\mathcal{L}_2)-\Re(\mathcal{C})(\mathcal{A}_1-\mathcal{A}_2)-\Im(\mathcal{C})(\mathcal{L}_1-\mathcal{L}_2),
\end{equation}
defined in terms of the Lorentzian~$\mathcal{L}$ and
dispersive~$\mathcal{A}$ functions that characterise the emission of
the lower (1) and upper (2) eigenstates (\emph{dressed states}):
\begin{align}
  \mathcal{L}_{\substack{1\\2}}(\omega)=&\frac{1}{2\pi}\frac{{\Gamma_+\pm
      \Im(R)}}{({\Gamma_+\pm \Im(R)})^2+(\omega-(\omega_a{-\frac{\Delta}{2}\mp \Re(R)}))^2}\,,\label{eq:ThuMar6002803UTC2008}\\
  \mathcal{A}_{\substack{1\\2}}(\omega)=&\frac{1}{2\pi}\frac{\omega-(\omega_a{-\frac{\Delta}{2}\mp
      \Re(R)})}{({\Gamma_+\pm
      \Im(R)})^2+(\omega-(\omega_a{-\frac{\Delta}{2}\mp
      \Re(R)}))^2}\,,\label{eq:ThuMar6153238UTC2008}
\end{align}
where the complex coefficient~$\mathcal{C}$ is defined as:
\begin{equation}
  \label{eq:ThuMar6001344UTC2008}
  \mathcal{C}=\frac{1}{R}\left[\Gamma_-+i\frac{\Delta}{2}+\frac{i\frac{g^2}2(\gamma_aP_b-\gamma_bP_a)(i\Gamma_+-\frac{\Delta}2)}
    {g^2\Gamma_+(P_a+P_b)+P_a\Gamma_b(\Gamma_+^2+(\frac{\Delta}{2})^2))}\right]
\end{equation}
and the complex \emph{Rabi splitting} as:
\begin{equation}
  \label{eq:ThuMar6125254UTC2008}
  R=\sqrt{g^2-\Big(\Gamma_-+i\frac{\Delta}{2}\Big)^2}\,.
\end{equation}

\begin{figure}[hbtp]
  \centering
  \includegraphics[width=\linewidth]{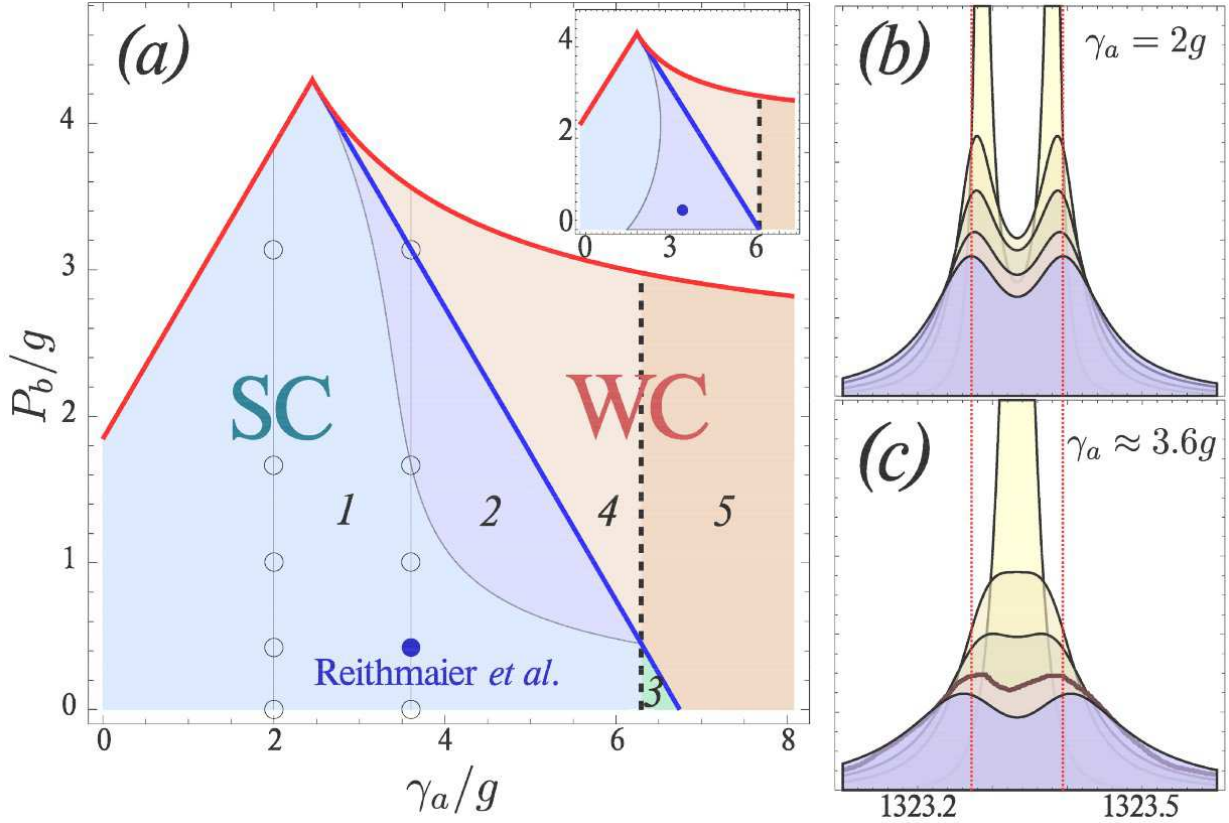}
  \caption{$(a)$ Regions of strong (blue) and weak (red) coupling at
    resonance in the space of parameters~$(\gamma_a, P_b)$, with
    $\gamma_b\approx2.3g$ and~$P_a\approx0.12\gamma_a$ fitting the
    experiment of Reithmaier \emph{et al.}~\cite{reithmaier04a},
    marked by a plain blue point.  Region~1 exhibits line-splitting,
    while in the darker area~$2$, although still in SC, the splitting
    cannot be resolved.  The dashed vertical line marks the criterion
    for SC in absence of pumping, giving rise to region~3 where SC is
    recovered (with line-splitting) thanks to pumping and region~4
    where it is lost because of it. In inset, the same but
    for~$P_a=0$, in which case the line-splitting
    of~\cite{reithmaier04a} would not be resolved.  $(b)$ and~$(c)$:
    Spectra of emission with increasing exciton pumping~$P_b$ marked
    by the hollow points in panel~$(a)$. For~$\gamma_a=2g$
    in~$(b)$, SC is retained throughout and made more visible.
    For the best fit parameter, $\gamma_a=3.6g$ in~$(c)$,
    line-splitting is lost increasing pumping, first because the
    line-splitting is not resolved (region~2 of~$(a)$) then because
    the system goes into WC (region~4 of~$(a)$).}
  \label{ThuMar6132015UTC2008}
\end{figure}

The spectral shape given by
formulas~(\ref{eq:ThuMar6001103UTC2008}--\ref{eq:ThuMar6125254UTC2008})
is that of a coupled (SC and WC) system in a steady state maintained
by incoherent pumping with rates~$P_{a,b}$, with the QD excitation
obeying Bose statistics. This is a valid approximation when the QD is
large (it becomes exact in the limit of a quantum well) or in any case
when the number of excitations is vanishing. If the QD follows Fermi
statistics, the analytical expression for the spectrum is lost. We
therefore keep this case out of the present discussion.  In the case
of Ref.~\cite{reithmaier04a}, both large QDs and low excitations were
used, and we verified numerically that a Fermionic model is less
appropriate. The basic structure of
eqn~(\ref{eq:ThuMar6001103UTC2008}) is the same as in other
descriptions of SC, such as the decay of an initial state.  Namely it
consists of two peaks, each the sum of a Lorentzian and of a
dispersive part. The limit of vanishing pumping is formally equal to
the particular case of the decay of an initial state with independent
initial populations, whose ratio is the same as that of the pumping
rates. The main result is to be found in the way incoherent pumping,
even if it is small, affects this intrinsic structure of SC through
eqns~(\ref{eq:ThuMar6001344UTC2008}--\ref{eq:ThuMar6125254UTC2008}).
This is demonstrated by confronting the theory with the experiment. In
Fig.~\ref{ThuMar6011242UTC2008}, we have optimised the \emph{global}
nonlinear fit of the results from Ref.~\cite{reithmaier04a} with
eqns~(\ref{eq:ThuMar6001103UTC2008}--\ref{eq:ThuMar6125254UTC2008}).
That is, the detuning ($\omega_a$ and~$\omega_b$) and pumping rates
($P_a$ and~$P_b$) are the fitting parameter from one curve to the
other, while~$g$ and $\gamma_{a,b}$ have been optimised but kept
constant for all curves. We find an excellent overall agreement, that
instructs on many hidden details of the experiment. 

First, the model provides more reliable estimations of the fitting
parameters than a direct reading of the line-splitting at resonance or
of the linewidths far from resonance: The best-fitting coupling
constant is $g=61\mu$eV. The value for~$\gamma_a=220\mu$eV is
consistent with the experiment (the authors place it at $180\mu$eV but
from a Lorentzian fit of the 5K curve in the assumption that the
system is not strongly-coupled here, where our model shows this to be
a poor approximation), and the value for~$\gamma_b$, that is the most
difficult to estimate experimentally, is reasonable in the assumption
of large QDs, as is the case of those that have been used to benefit
from their large coupling strength.  Our point here is not to conduct
an accurate statistical analysis of this particular work but to show
the excellent agreement that is afforded by our model with one of the
paradigmatic experiment in the field.  
Such a good global fit cannot be obtained without taking into account
the effect of pumping, even when it is small. More interestingly, it
is necessary to include both the exciton pumping~$P_b$ (expected from
the experimental configuration) but also the cavity pumping~$P_a$.
The latter requirement comes from the fact that in such samples, there
are numerous QDs weakly-coupled to the cavity in addition of the one
that undergoes SC.  Beyond this QD of interest, a whole population of
``spectator'' dots contributes an effective cavity pumping, which
looms up in the model as a nonzero~$P_a$. The fitting pumping rates
vary slightly with detuning, as can be explained by the change in the
effective coupling of both the strongly-coupled dot with the cavity
(pumping tends to increase out of resonance) and the spectator QDs
that drift in energy with detuning.  We find as best fit parameters at
resonance~$P_a\approx0.12\gamma_a$ and~$P_b\approx0.18\gamma_b$ (the
mean over all curves is $\bar P_a\approx0.15\gamma_a$ and~$\bar
P_b\approx0.28\gamma_b$ with rms deviations of~$\approx10\%$). The
existence of~$P_a$ in an experiment with electronic pumping is
supported by the authors of \cite{reithmaier04a} who observed a strong
cavity emission with no QD at resonance.  We shall see in the
following the considerable importance of this fact to explain the
success of their experiment.

From a fundamental point of view, our incoherent pumping model of SC
not only fills in a gap in extending the theory to the steady-state
case where the excitation is not given (sometimes arbitrarily) as an
initial state, it also defines new criteria for SC.  The conventional
one, from the condition that~$R$ be real at resonance, is, neglecting
pumping:
\begin{equation}
  \label{eq:ThuMar6161733UTC2008}
  g>|\gamma_-|\,.
\end{equation}
With incoherent pumping, it becomes:
\begin{equation}
  \label{eq:WedMar12145005UTC2008}
  g>|\Gamma_-|\,.
\end{equation}
The full extent of this new criterion can be appreciated in
Fig.~\ref{ThuMar6132015UTC2008}, where is displayed in shades of blue
the region where~$R$ is real and in shades of red where it is pure
imaginary. This corresponds respectively to oscillations or not in the
time correlators and therefore to oscillations (SC) or damping (WC) of
the fields. In white (delimited by the red frontier) is the region
where there is no steady state because of a too-high pumping. The
dashed black line delimits the conventional (without incoherent
pumping) criterion, eqn~(\ref{eq:ThuMar6161733UTC2008}). Regions $3$
and~$4$ show how pumping can make a qualitative difference. In
region~$3$, given by $|\Gamma_-|<g<|\gamma_-|$, SC is not expected
according to eqn~(\ref{eq:ThuMar6161733UTC2008}), but holds thanks to
pumping, eqn~(\ref{eq:WedMar12145005UTC2008}) (in this case, thanks to
cavity pumping~$P_a$; in inset, $P_a$ is set to zero and this region
has disappeared). In region~$4$, given by $|\gamma_-|<g<|\Gamma_-|$,
where on the contrary SC is expected according to
eqn~(\ref{eq:ThuMar6161733UTC2008}), it is lost because of pumping. In
the regions~$1$ and~$5$, the effect of the pump is quantitative only,
renormalising the broadening and splitting of the peaks, but is still
important to provide a numerical agreement with experimental data.
Region $2$ is that where, although in SC, only one peak is observed in
the PL spectrum because of the broadening of each peak being too
important as compared to their splitting. The position where we
estimate the result of Ref.~\cite{reithmaier04a} in this diagram
validates that SC has indeed been observed in this experiment.  In
inset, however, one sees that in the case where the cavity
pumping~$P_a$ is set to zero keeping all other parameters the same,
the point falls in the dark region~$2$ where, although still in SC,
the line-splitting cannot be resolved.  Even if it is possible in
principle to demonstrate SC through a finer analysis of the crossing
of the lines, it is obviously less appealing than a demonstration of
their anticrossing.  This is despite the fact that the case of~$P_a=0$
is equally, if not more, relevant as far as SC is concerned, as it
corresponds to the case where only the QD is excited, whereas in the
case of Fig.~\ref{ThuMar6011242UTC2008}, it also relies on cavity
photons.  With the populations involved in the case of the best fit
parameters that we propose---$n_a\approx0.15$ from
eqn~(\ref{eq:FriOct26003321UTC2007})---one can still read in
Reithmaier \emph{et al.}'s experiment a good \emph{vacuum} Rabi
splitting, so the appearance of the line-splitting with~$P_a$ is not
due to the photon-field intensity.  Rather, the system is maintained
in a quantum state that is more photon-like in character, which is
more prone to display line splitting in the cavity emission with the
parameters of Fig.~\ref{ThuMar6011242UTC2008}. One can indeed check
that the PL spectrum without pumping for a strongly-coupled state
prepared as a photon exhibits a line-splitting whereas the same system
prepared as an exciton does not show it. This is the same principle
that applies here, with the nature of the state (photon-like or
exciton-like) resolved self-consistently by pumping.  In this sense,
there is indeed an element of chance involved in the SC observation,
as one sample can fall in or out of region~2 depending on whether or
not the pumping scheme is forcing photon-like states.

A natural experiment to build upon our results is to tune pumping.  In
our interpretation, it is straightforward experimentally to
change~$P_b$, but it is not clear how~$P_a$ would then be affected, as
it is due to the influence of the crowd of spectator QDs, not directly
involved in SC.  In Fig.~\ref{ThuMar6132015UTC2008}$(c)$, we
hold~$P_a$ to its best fit parameters and vary~$P_b$ in the best fit
case~$\gamma_a=3.6g$ on panel $(c)$, then for~$\gamma_a=2g$ on
panel~$(b)$, where the system is in SC for all possible values
of~$P_b$. Spectra are displayed for the values of~$P_b$ marked by the
points in~$(a)$.  Two very different behaviours are observed for two
systems varying slightly in one of their parameters. In one case
($b$), strong renormalisation of the linewidths and splitting results
from Bose effects in a system that retains SC throughout. In the other
($c$), line-splitting is lost and transition towards WC then follows.
At very high pumping, the model breaks down.  The most interesting
possibility is that the QD becomes Fermi-like, in which case the
annihilation operator~$b$ should be a two-level system.  Loss of
line-splitting also results in this case with a smaller decrease in
the linewidth than is observed here at moderate pumpings and a
subsequent increase at higher pumpings, as the system reaches the
self-quenching regime (when~$b$ is a Bose operator, linewidths tend to
zero as the populations diverge when effects assured to take place at
high density, such as particles interaction, are neglected).  A
careful study of pump-dependent PL can tell much about the underlying
statistics of the excitons and the precise location of one experiment
in the SC diagram.

In conclusion, we obtained a self-consistent analytical expression for
the spectrum of emission of a coupled light-matter system in a steady
state maintained by a continuous incoherent pumping.  Our formalism
fully takes into account the effects of the incoherent pump, that, by
randomising the arrival time of the excitation, averages out the Rabi
oscillations and through the interplay of pumping and decay rates,
imposes a quantum steady state that influences considerably the
observed spectra. Close to its threshold, SC can be lost or imposed by
increasing pumping, and even when SC holds, the decoherence can hinder
line-splitting.  Prospects for applications of SC with semiconductor
heterostructures are great, provided that a quantitative understanding
of the system can guide the advances now that the qualitative effects
have been observed. Our model offers such a fine theoretical
description, while still staying at a fundamental level with a
transparent physical interpretation. We showed that the experimental
reports of SC can be fully explained taking into account conjectures
such as the influence of weakly-coupled dots, and other similar
factors that can be better taken advantage of in the future.

Authors are grateful to Dr.~Sanvitto and Dr.~Amo for helpful comments.
Support by the Spanish MEC under contracts QOIT Consolider
CSD2006-0019, MAT2005-01388 and NAN2004-09109-C04-3 and by CAM under
contract S-0505/ESP-0200 is acknowledged.

\bibliography{Sci,psip}

\end{document}